\documentclass[9pt,twocolumn,twoside]{osajnl}

\journal{josab} 
\usepackage{pifont}

\setboolean{shortarticle}{false}

\title{Spatial Fano resonance of a dielectric microsphere impinged by a Bessel beam}

\author[1,*]{V.~Klimov}
\author[2,\ding{61}]{R.~Heydarian}
\author[2,3]{C.~Simovski}

\affil[1]{Department of Optics, Lebedev Physical Institute, Russian Academy of Sciences, 53, Leninsky Prospekt, 119991, Moscow, Russia}
\affil[2]{Department of Electronics and Nano-Engineering, Aalto University, P.O. Box 15500, FI-00076 Aalto, Finland}
\affil[3]{Faculty of Physics and Engineering, ITMO University, 199034, Birzhevaya line 16, Saint-Petersburg, Russia}

\affil[*]{Corresponding author: klimov256@gmail.com}
\affil[\ding{61}]{Corresponding author: reza.heydarian@aalto.fi}

\begin{abstract}
General concept of Fano resonance is considered so that to show the possibility of this resonance in space. Using a recently found solution for a Bessel wave beam impinging a dielectric sphere, we analyze the electromagnetic fields near {a microsphere with} different optical size and permittivity values. We theoretically reveal a spatial Fano resonance when a resonant mode of the sphere interferes with {an amount of } non-resonant modes. This resonance results in a giant jump of the electric field behind the sphere impinged by the first-order Bessel beam.  The local minimum of the electromagnetic field turns out to be  noticeably distanced from the rear edge of the microsphere. However, this is a near-field effect and we prove it.  We also show that this effect can be utilized for engineering a submicron optical trap with unusual and useful properties.
\end{abstract}

\setboolean{displaycopyright}{false}

\begin{document}

\maketitle

\section{Introduction}

One of the main challenges of nano-optics is to produce nanostructured light {which can have spatial features well} below the wavelength. Such nanostructured light is useful for applications such as microscopy \cite{1,2,3} and optical tweezers \cite{4,5}. Recently deep 3D subwavelength focusing was achieved with making use of superoscillating superposition of Bessel beams \cite{6}.
In the optical literature the Fano resonance initially revealed in quantum mechanics was discussed for the frequency spectrum of the scattered or emitted light (see e.g. in \cite{limonov7,fano8,fano9}). In general, the Fano resonance occurs when a discrete quantum/optical state interferes with a continuum band of states, and it is manifested in the absorption spectrum, $\sigma(E)$, with the shape described by the famous Fano formula: 

\begin{equation}
    \sigma(E)=D^2 \frac{(\Omega+q)^2}{\Omega^2+1}
    \label{eq1}
\end{equation}

\noindent
where $E$ is the energy, $q=\cot{\delta}$ is the Fano parameter, $\delta$ is the phase shift of the continuum, $\Omega=2(E-E_0)/\Gamma$, where $\Gamma$ and $E_0$ are the resonance width and energy, respectively, and $D^2=4\sin^2\delta$ (in the form presented in \cite{connerade10}). 

\begin{figure}[t!]
\centering
\includegraphics[width=0.4\textwidth,height=0.25\textwidth]{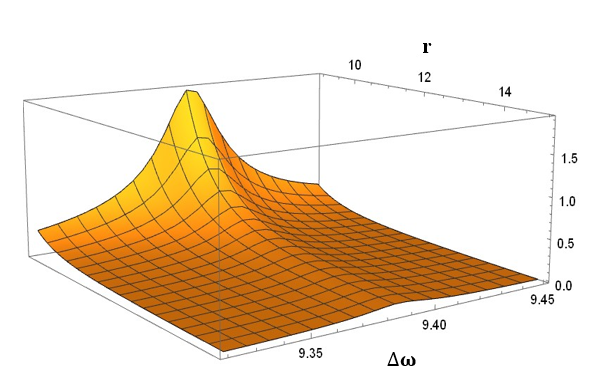}
\caption{Resonant mode spectrum using the example of $TM_{10,0,1}$ Mie resonance $E(r,\omega)=q_{Mie,10}(\omega)H_{10}^{(1)}(kr)/kr$  
}
\label{fig1}
\end{figure}

In near-field optics, however, spatial modes interfere, and the quantity of interest to us is the spatial distribution of intensity at a given frequency, which in the case of two modes has the form:
\begin{equation}
    I(r,\omega)=|E_1(r,\omega)+E_2(r,\omega)|^2.
    \label{eq2}
\end{equation}
Here $r$ is a coordinate, for example the length of a radius vector with the coordinate origin at the center of a sphere 
experiencing the diffraction. The intensity structure of each of the spherical modes has the same unimodal form, shown in Fig. \ref{fig1} as a function of both $r$ and $\omega$.

\noindent
Figure \ref{fig1} shows that {along both frequency and spatial axes,} there is a maximum of the mode intensity. On the frequency axis, the resonance has Lorentzian shape; on the {spatial} axis, the field decreases more rapidly.
When two as in (\ref{eq2}) or more modes are excited, interference effects {with} constructive {or} destructive nature are possible. At a fixed spatial position of the observation point, the usual Fano resonance arises as a function of frequency. At a fixed frequency, there is interference of spatial modes, and with destructive interference, spatial Fano resonance can occur. It is this effect and its applications that this work is devoted to. Here we will demonstrate the possibility to create a  nanostructured light using a spatial Fano resonance produced by a dielectric sphere impinged by a symmetric Bessel beam \cite{7,8}.

Making use an axially symmetric Bessel beam allows one to excite only $TM$ or $TE$ modes and to prevent their interference. Moreover, the presence of an additional parameter $\beta$ (so-called conical angle of the radially polarized beam) allows one to control the interaction of the light beam with specific $TM$ or $TE$ modes in wide limits.  

Using this solution in \cite{8} the so-called pseudo-modes (resonant forced solutions) with practically unlimited radiative quality factors were revealed for a dielectric sphere with a moderate optical size {$x_0\equiv k_0a\sim1$ (where $k_0$ is wavenumber of free space and $a$ is the sphere radius)} and high permittivity $\varepsilon$. At their resonance frequencies these field patterns do not leak, and the scattering is practically absent. Unlike bound states in the continuum (for which the last property had been known), pseudo-modes cannot exist in a lossless sphere without an external excitation. Moreover, they cannot be excited by a plane wave. A set of few Bessel beams is needed. However, this result is related to the subwavelength field concentration inside the sphere, and can be, therefore, treated as a new but not very novel near-field effect.

\begin{figure}[h!]
\centering
\includegraphics[width=0.45\textwidth,height=0.3\textwidth]{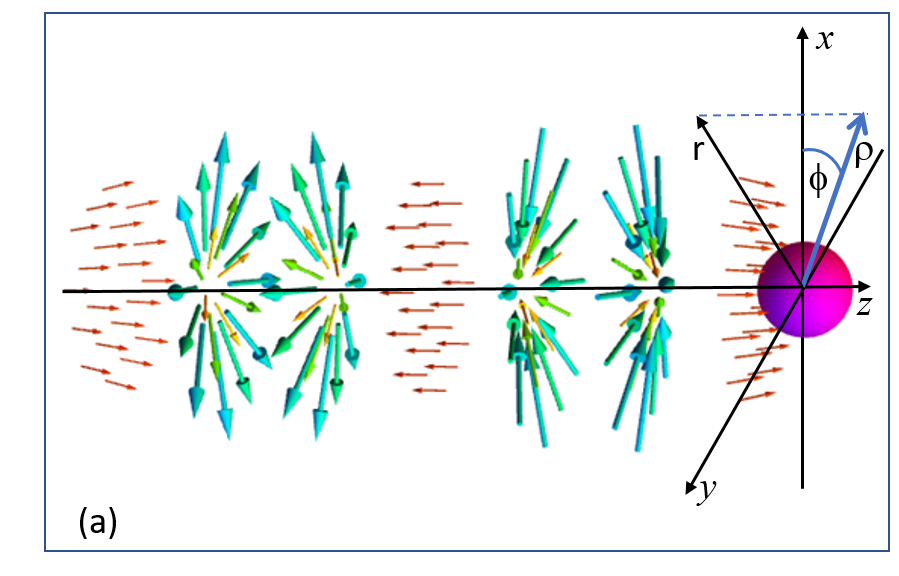}
\includegraphics[width=0.4\textwidth,,height=0.3\textwidth]{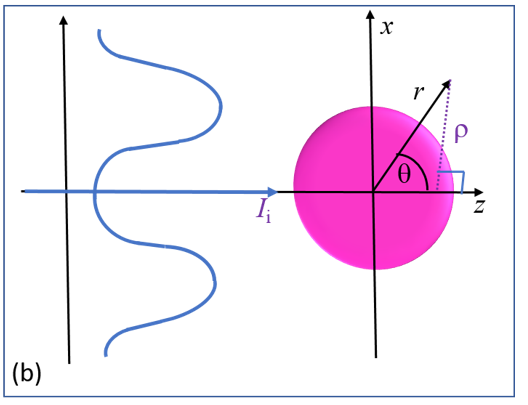}
\caption{ (a). Simulated vector diagram of electric field of the incident Bessel beam impinging the microsphere in paraxial domain.
 (b). Electric field intensity distribution $I_i=({E^i_{\rho})}^2+{(E^i_z)}^2$ of the incident beam shown qualitatively in the plane $(x-z)$. Initial Cartesian frame is shown as well as cylindrical and spherical coordinate systems.  
}

\label{fig2}
\end{figure}

In the present paper, we reveal a very novel type of a near-field effect on an example of a dielectric sphere impinged by an axially symmetric Bessel beam. Our spatial Fano resonance corresponds to {the strong suppression (practically up to zero) and strong enhancement (by two order of magnitude)} of the local electric field in the free-space area behind the sphere. These properties make the revealed near-field effect very unusual. {Instead of a strongly subwavelength spots in which the local intensity is drastically enhanced,} we observe a strong electric field maximum at the sphere rear edge and practically null electromagnetic field at a point distant from the rear edge of the sphere by $(0.3-0.5)\lambda$. Behind this sharp local minimum we observe in our calculations a spread local maximum.  

To observe our effect, a single incident Bessel beam and the modest permittivity of the sphere, inherent to 
different types of optical glass (whose $\varepsilon$ may have values in the range $2.7-3.2$) or transparent metal oxides/sulfates/sulfides (permittivities in the interval  $\varepsilon=3-4$) are sufficient. The sphere of such the material {must be optically large ($x_0\gg 1$). For given radius {$a$} $x_0$ can be considered as the normalized frequency, i.e. our effect is not observed at low frequencies.} Another necessary condition of the effect is smallness of the beam conical angle {$\beta$}. The incident beam intensity should have the local minimum on the beam axis (where the electric field is polarized along $z$ and the magnetic field is identically zero) so that the sphere radius is smaller than the radius of the first beam intensity maximum. This excitation of the sphere is illustrated by Fig.~\ref{fig2}. 

The outline for the rest of the article is as follows. In Section \ref{sec2}, using the example of Bessel beam scattering by a dielectric sphere, we will show that the spatial Fano resonance can arise in such a system and the conditions for its occurrence are found. In Section \ref{sec3}, we explore the possibility of using spatial Fano resonances for creation of efficient 3D optical traps for nanoparticles and even of submicron optical traps suitable for a long trapping of cold atoms and molecules.

\section{Spatial Fano Resonance in the presence of a dielectric microsphere impinged by a Bessel beam}
\label{sec2}

Let a Bessel beam of the first order with azimuthally polarized magnetic field (then the electric field has both radial $\rho$ and axial $z$ components) symmetrically impinges a dielectric sphere of permittivity $\varepsilon$ and radius {$a$} as it is shown in Fig.~\ref{fig2}. {The wave vector components of the Bessel beam form a cone having a conical angle $\beta$ with the $z$ axis. For a sphere whose its center is placed at the origin, we may write the  incident fields for such Bessel beam in the cylindrical coordinates $(\rho,\varphi,z)$ as:}
\begin{eqnarray}
\label{incident}
    H_{\varphi}^{in}=H_0J_1(k_0\rho\sin\beta)e^{i(k_0z\cos\beta)} \nonumber \label{eq3} \\ 
    E_{\rho}^{in}={\eta_0H_0}\cos\beta J_1(k_0\rho\sin\beta)e^{i(k_0z\cos\beta)}  \\ 
    E_{z}^{in}= i{\eta_0H_0} {\sin{\beta}} J_0(k_0\rho\sin\beta)e^{i(k_0z\cos\beta)}, \nonumber
\end{eqnarray}

where $J_{0,1}$ are cylindrical Bessel functions and $\eta_0$ is the free-space wave impedance. {Here  we  adopt  the  time dependence  $exp(-i\omega t)$.} Let us note that on the beam axis {($z$ axis)} both $H_{\varphi}^{in}$ and $E_{\rho}^{in}$ are equal to zero. 
In \cite{8} it was shown that the expansion of Bessel beam in spherical harmonics has the following form:

\begin{eqnarray}
    H_{\varphi}^{in}(r,\theta)=-iH_0\sum_{n=1}^{\infty}E_n j_n(k_0r)\Psi_n(\theta)\Psi_n(\beta), \nonumber \label{eq4}  \\ 
    E_{r}^{in}(r,\theta)=\frac{{\eta_0H_0}}{k_0r}\sum_{n=1}^{\infty}i^n(2n+1) j_n(k_0r)P_n(\cos{\theta})\Psi_n(\beta),  \\ 
    E_{\theta}^{in}(r,\theta)=-\frac{{\eta_0H_0}}{k_0r}\sum_{n=1}^{\infty}E_n [Rj_n(R)]_{R=k_0r}^{'}\Psi_n(\theta)\Psi_n(\beta). \nonumber
\end{eqnarray}
\noindent
{Here $E_n=i^n{(2n+1)}/{n(n+1)}$,  $\Psi_n(\theta)={dP_n(\cos{\theta})}/{d\theta}$, $^{'}\equiv{\partial}/{\partial R}$, $j_n(R)$ and $P_n(\cos{\theta})$ are the spherical Bessel functions and the Legendre polynomials, respectively.}  Using (\ref{eq4}), scattered fields are presented as follows:

\begin{eqnarray}
    H_{\varphi}^{sc}(r,\theta)=-iH_0\sum_{n=1}^{\infty}a_nE_n h_n^{(1)}(k_0r)\Psi_n(\theta)\Psi_n(\beta), \nonumber \label{eq5}\\ 
    E_{r}^{sc}(r,\theta)=\frac{H_0{\eta_0}}{k_0r}\sum_{n=1}^{\infty}a_ni^n(2n+1) h_n^{(1)}(k_0r)P_n(\cos{\theta})\Psi_n(\beta),  \\ 
    E_{\theta}^{sc}(r,\theta)=-\frac{H_0{\eta_0}}{k_0r}\sum_{n=1}^{\infty}a_nE_n [Rh_n^{(1)}(R)]_{R=k_0r}^{'}\Psi_n(\theta)\Psi_n(\beta). \nonumber
\end{eqnarray}
\noindent
Corresponding expansions can be written for the internal electric and magnetic fields. Equating the $\varphi$ and $\theta$ components of the internal and external fields at $r=a$, we find the Mie coefficients in the form:

\begin{equation}
    a_n=-\frac{\varepsilon j_n(R_1)[R_0j_n(R_0)]'-j_n(R_0)[R_1j_n(R_1)]'}{\varepsilon j_n(R_1)[R_0h_n^{(1)}(R_0)]'-h_n^{(1)}(R_0)[R_1j_n(R_1)]'},  
    \label{eq7}
\end{equation}
\noindent
{in which it is denoted $R_{0,1}=ak_{0,1}$, $k_1=k_0\sqrt{\varepsilon}$.}

\begin{figure}[h!]
\centering
\includegraphics[width=0.49\textwidth,height=0.35\textwidth]{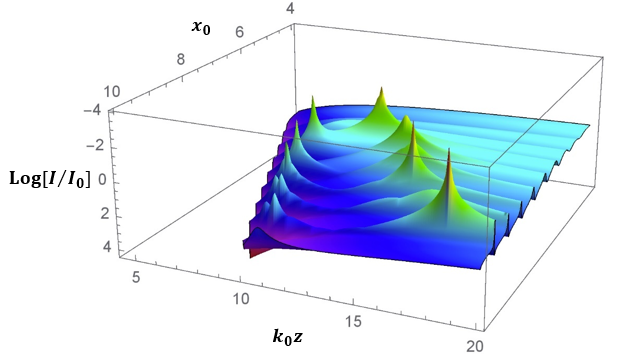}
\caption{{Normalized electric intensity} versus the size parameter $x_0$ and the normalized $z$ coordinate for the case $\varepsilon=3.1$ and $\beta=0.01$ (weakly focused Bessel beam). Sharp maxima in this plot are in fact very sharp minima due to the inversion of the ordinate and its logarithmic scale.}
\label{fig3}
\end{figure}

Using this solution we have calculated the {normalized electric intensity on the axis $z$ ($\theta=0$)} behind the sphere as a function of $z$ and $x_0$ for $\varepsilon=3.1$ and $\beta=0.01$. The result of these calculations is shown in Fig.\ref{fig3}.

\begin{figure}[h!]
\centering
\includegraphics[width=0.44\textwidth,height=0.3\textwidth]{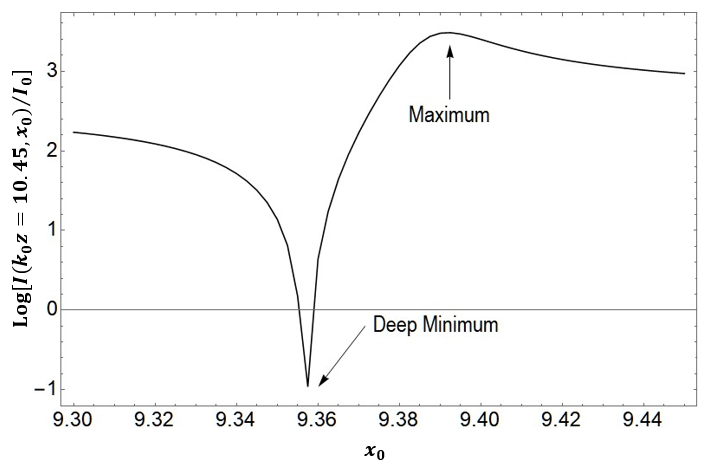}
\caption{Ordinary Fano resonance $\varepsilon=3.1$ and $\beta=0.01$ at the point $k_0z=10.45$.   
}
\label{fig4}
\end{figure}

{Fig. \ref{fig3}} clearly shows deep minima (for better visibility shown as maxima due to the inversion of the ordinate), which correspond to ordinary (spectral) Fano resonances (when $z=$ const and $x_0$ varies) or spatial Fano resonances (when $x_0=$ const and $z$ varies). Let us consider in more detail the Fano resonance near the point $x_0=9.36$ and {$k_0z=10.45$}. Figure \ref{fig4} shows the ordinary Fano resonance for these parameters (note that at the $z$ axis $E=E_z$ and the electric intensity is $I=E_z^2$). 

\begin{figure}[h!]
\centering
\includegraphics[width=0.4\textwidth,height=0.25\textwidth]{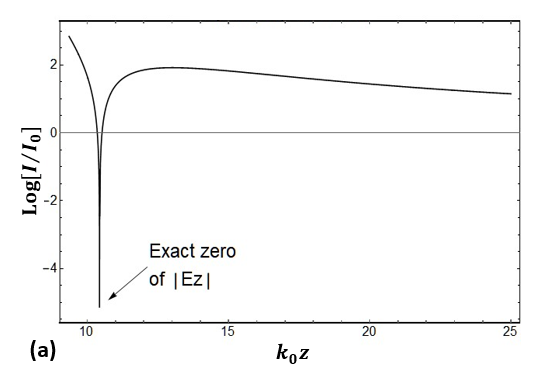}
\includegraphics[width=0.4\textwidth,height=0.3\textwidth]{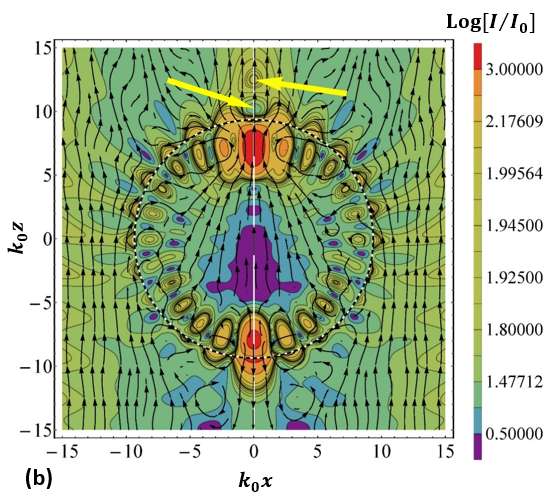}
\includegraphics[width=0.4\textwidth,height=0.3\textwidth]{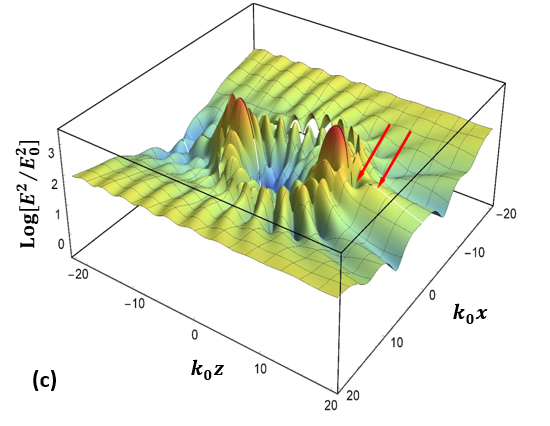}
\caption{Spatial Fano resonance for the case $\varepsilon=3.1$ and $\beta=0.01$ when the size parameter $x_0=9.3568$ is selected so that to obtain an exact zero of the field. Coordinate dependency of the normalized electric intensity (on the beam axis behind the sphere) (a) Color map of the electric intensity (log scale) combined with the map of the local Poynting vector (b). Two yellow arrows point out the local maximum and the local minimum. A 3D plot of the intensity in the log scale is added for the better visibility of the effect (c). Two red arrows correspond to the same points. 
}
\label{fig5}
\end{figure}

 A remarkable point is that this dependence of the electric field on the size parameter leads to the similar dependence of the spatial coordinates. The spatial Fano resonance  is visualized in three plots of Fig. \ref{fig5}. In Fig. \ref{fig5}(a) we show the normalized electric intensity $(E_z^2+E_\rho^2)/I_0$, where {$I_0=(\eta_0H_0)^2$} along the beam axis $z$. In Fig. \ref{fig5}(b) we show both the normalized electric intensity and the local Poynting vector in the ($x-z$) plane. Due to the problem symmetry there is no $\varphi$ dependence {and $x$ is an arbitrary axis orthogonal to $z$.} These plots correspond to $x_0=9.3568$, $\varepsilon=3.1$ and $\beta=0.01$. 
 The color bar here is also chosen in the log scale. A 3D plot of the intensity (also in the log scale) is added in Fig. \ref{fig5}(c) for the better visibility of the effect.

 The high values $E_z^2/E_0^2\sim 200$ at the rear edge of the sphere  $k_0z=10$ can be explained by the multipole resonance. 
 Our sphere is rather optically large and our wave beam efficiently excites a large set of multipoles. Any frequency in the range $x_0\sim 10$ can be attributed to one of the resonance bands of these multipoles. Therefore, the hot spots of the electromagnetic field inside such the sphere are expected. In our azimuthally symmetric problem one of these hot spots always  covers the rear edge $(\rho=0,\, z={a})$ of the sphere.However, besides of the enhancement of $E_z$ at this rear point, we also see a noticeable enhancement of the local electric intensity in the region noticeably shifted behind the sphere.
 
\begin{figure}[htbp!]
\centering
\includegraphics[width=0.44\textwidth,,height=0.3\textwidth]{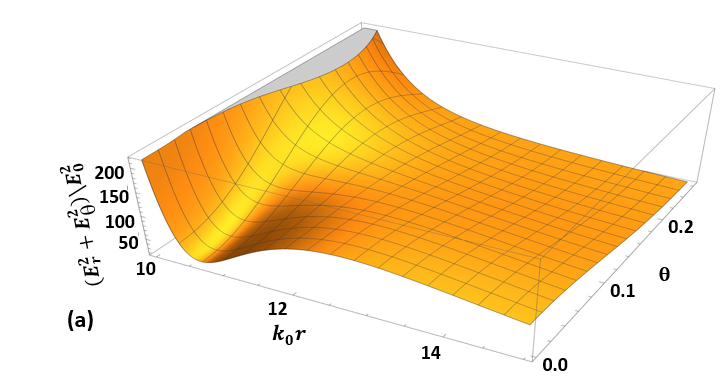}
\includegraphics[width=0.44\textwidth,,height=0.26\textwidth]{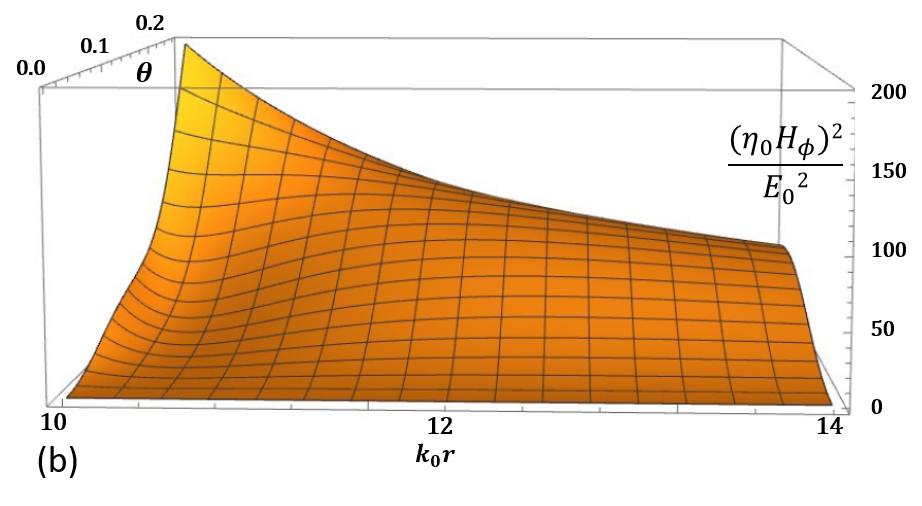}
\caption{(a) Electric intensity normalized to $E_0^2$ in the area behind the sphere in spherical coordinates. 
(b) Magnetic intensity plot normalized to $(E_0^2/\eta_0^2)$ in the same area. Fixed parameters are as follows: $x_0=9.348$, $\beta=0.01$ and $\varepsilon=3.1$.}
\label{fig6}
\end{figure}

 Inside the sphere with the size parameter $x_0\approx 10$ we see 24 intensity maxima (of different magnitude) located around near the sphere surface. It corresponds to $12$ wavelengths of the standing creeping wave. In the language of the cavity resonances, this pattern corresponds to the resonance of the mode with $n=12$. More exactly, this is the transverse magnetic mode $TM_{12,0,1}$, where $n=12$ is orbital mode number, $m=0$ is azimuthal mode number and $p=1$ is the radial mode number, respectively (see e.g. in \cite{9}). Since any high-order Mie resonance is the same as the excitation of a whispering gallery mode whose wavelength in the sphere medium $\lambda_{\rm m}=\lambda/\sqrt{\varepsilon}$ can be found as $\lambda_{\rm m}=2\pi {a}/n$, for the mode $n=12$ the sphere perimeter should cover namely $12$ wavelengths $\lambda_{\rm m}$ ($24$ intensity maxima around the sphere).
 Fig. \ref{fig6}(a) shows the normalized electric intensity  $(E_r^2+E_\theta^2)/E_0^2$  for the case $x_0=9.348$, $\varepsilon=3.1$ as a function of both $r$ and $\theta$ in the spherical coordinates system ($r$, $\theta$, $\varphi$) centering the sphere. The figure shows that the Fano minimum may be engineered very sharp -- the gradients of the electric field around it are very high. This feature of the effect can be used for the trapping of nanoparticles and, perhaps, even atoms/molecules. Fig. \ref{fig6}(d) depicts the normalized magnetic intensity as the function of both spherical coordinates $r$ and $\theta$ (recall that the problem has axial symmetry). We see on this figure only the tail of the rear internal hot spot of the 12-th mode. It means that strong concentration on axis $z$ behind the sphere holds only for the longitudinal electric field, nothing similar occurs for the magnetic field. 
 
 Let us now discuss in details the unusual interference effect underlying our Fano resonance.  
 The mode shown in Fig. \ref{fig5}(b), though resonant, does not completely dominate over the field created by all other terms of the Mie expansion. Its field visibly interferes with that of the infinite amount of all other modes. Just behind the sphere this interference is destructive and we see a sharp minimum{. A bit farther the interference becomes constructive and we see a maximum. Extrema of these constructive and destructive interference are marked by yellow and red arrows in Figs. \ref{fig5}(b) and \ref{fig5}(c),respectively. We guess that the amplification and vanishing of $E_z^2$ at points which are substantially distanced from the sphere correspond to the spatial Fano resonance.} This guess is confirmed by our further study. First, we analysis the large amount of Mie coefficients. {Contribution of the resonant one and of all the other ones, in terms of power, are practically equal} if $x_0$ corresponds to the sharp minima depicted in Fig. \ref{fig3}. 

\begin{figure}[h!]

\includegraphics[width=0.44\textwidth,,height=0.3\textwidth]{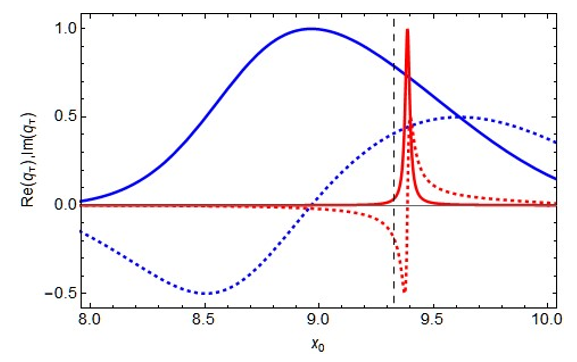}

\caption{Real and imaginary parts of $q_8$ (blue) and $q_{12}$ (red) for $\varepsilon=3.1$.}
\label{fig7}
\end{figure}

Indeed, the exact solution (\ref{eq5}) allows us to present the scattered electric field phasor on the axis $z$ behind the sphere where it is polarized along $z$ in the following form:

\begin{eqnarray}
    E_{z}^{sc}(z>a,\theta=0)= -\frac{E_0}{2k_0z} \nonumber \\ \label{eq8}
    \times\left\{\underbrace{\sum_{n=1}^{N-1}a_n(x_0)\sigma(n)h_n^{(1)}(k_0z)}_{\text{quasi-continuum, arg>0}}+\underbrace{a_N(x_0)\sigma(N)h_N^{(1)}(k_0z)}_{\text{resonant term, arg<0}}+\Delta\right\} 
\end{eqnarray}

\noindent
where $\sigma(n)= i^n (2n+1)\Psi_n(\beta)$ and $a_n$ are our Mie coefficients (\ref{eq7}).
Let the normalized frequency $x_{0N}$ be that of the $N$-th TM-polarized mode resonance, and let $N\gg 1$. Then in 
Eq. (\ref{eq8}) the first term (in brackets) is a quasi-continuum of those modes whose resonances hold at lower frequencies, the second term is the resonant one and the third term (representing the contribution of the modes with $n>N$) is negligibly small. In fact, for a given $x_0$, the number of modes noticeably excited by our beam is limited: $n\leq N$.   
In the left half ($x_0<x_{0N}$) of the resonance range ($x_0\approx x_{0N}$) there is a normalized frequency $x_0$ at which the first and the second terms in {Eq.(\ref{eq8})} have exactly opposite phases and amplitudes of the same order. These amplitudes vary in space because these modes are mainly composed by evanescent spatial harmonics. At a point where these amplitudes are equal to one another, the first and the second term of {Eq. (\ref{eq8})} cancel out. This fact follows from the structure of the Mie coefficients, the real and imaginary parts of which are shown in Fig. \ref{fig7}.

 Moreover even two modes with overlapping resonance bands may interfere in this way. In Fig. \ref{fig7} it is seen that slightly at the {right} of the 12-th TM maximum (dashed vertical line), the mode $n=$8 and the mode $n=N=$12 have the close modal amplitudes and opposite phases. The 8-th mode has a very wide resonance band and the narrow-band 12-th mode forms in its resonance band an ordinary Fano resonance with it. However, we saw that the ordinary Fano resonance in our case is accompanied by the spatial one. 

\begin{figure}[h!]
\includegraphics[width=0.44\textwidth,,height=0.25\textwidth]{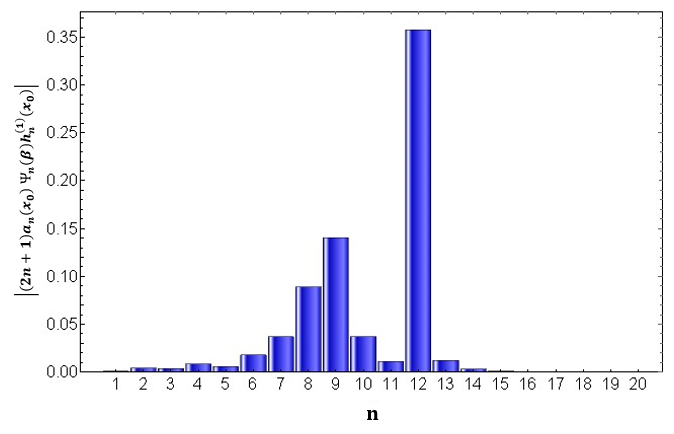}
\caption{Amplitudes of modes at the sphere rear edge ($kz=x_0$) for $\varepsilon=3.1$, $x_0=9.35679$, $\beta=0.01$. }
\label{fig8}
\end{figure}

In Fig. \ref{fig8} the amplitudes $(2n+1)\Psi_n(\beta)|a_nh_n^{(1)}(x_0)|$ of all modes at rear edge for $\varepsilon=3.1$, $x_0=9.35679$, $\beta=0.01$ are shown. In Fig. \ref{fig8} one can see the quasi-continuum of modes with $n\leq11$ and resonant term with $n=N=12$. Now the prerequisite of the Fano resonance is the combination of the opposite phases of the resonant mode and the quasi-continuum of lower modes and the same order of corresponding magnitudes.    

\begin{figure}[h!]

\includegraphics[width=0.5\textwidth,,height=0.2\textwidth]{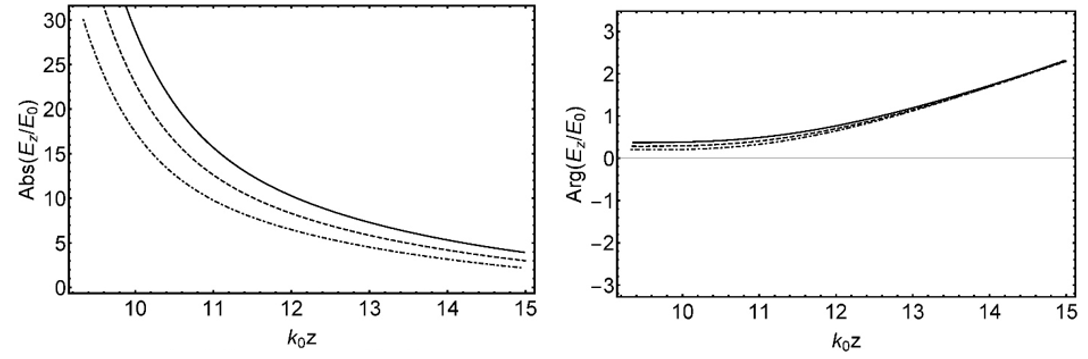}
\includegraphics[width=0.5\textwidth,,height=0.2\textwidth]{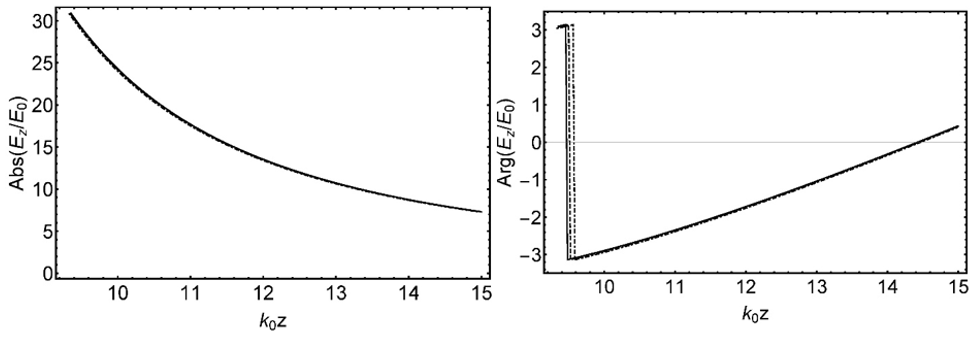}
\includegraphics[width=0.5\textwidth,,height=0.2\textwidth]{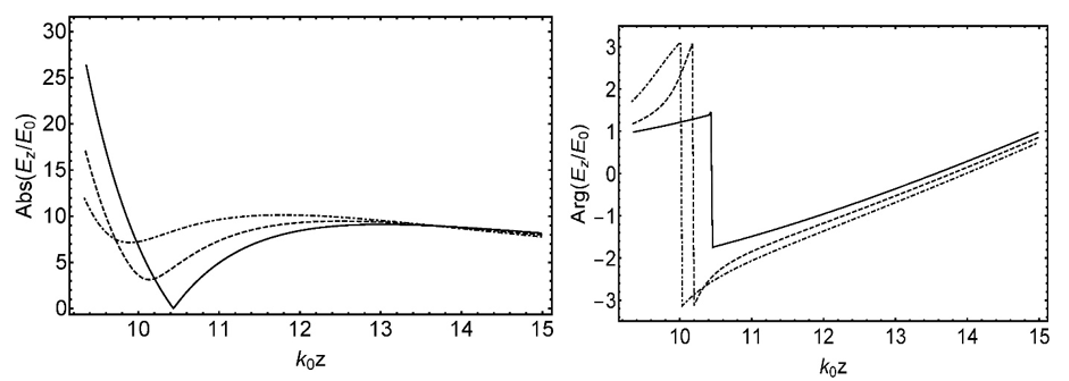}

\caption{Coordinate dependencies of the normalized electric intensity on the beam axis behind the sphere for three optical sizes of the sphere. Upper row -- the resonant mode amplitude and phase. Middle row -- the quasi-continuum amplitude and phase. Bottom row -- the amplitude and phase of the total field. Solid, dashed and dot-dashed curves correspond to $x_0=9.357$, 9.348, 9.335, respectively. In all cases $\beta=0.01$, $\varepsilon=3.1$.
}
\label{fig9}
\end{figure}

The formation of the Fano resonance due to this interference is illustrated by Fig. \ref{fig9} for the case, when  $\varepsilon=3.1$ and $x_0$ takes three values: $x_0=9.35679,\, x_0=9.348$ and $x_0=9.360$. In all three cases, the resonant mode is $N = 12$. Its field interferes with the sum of terms $n < N$ so that the maximum holds at the rear edge of the sphere and the minimum holds at a noticeable distance $\Delta z$ from the maximum. 
Top, middle and bottom panels of Fig. \ref{fig9} show it details how these field terms vary in space and interfere with one another. It is seen that the amplitudes decay with the distance from the sphere nearly exponentially because the evanescent waves dominate in both these field terms. It is also seen that for $x_0=9.335$, $\Delta z \approx\lambda/8$, for $x_0=9.348$, $\Delta z \approx\lambda/6$, and for $x_0=9.356787479$, $\Delta z \approx\lambda/4$. 
Clearly, this is a classical Fano resonance though it holds not in the spectral domain but in space and occurs at a noticeable distance from the rear edge $z=a$ of the sphere. It holds for an infinite set of frequencies from which we have shown three. It is also instructive to note an analogy between our Fano resonances and the well-known Wood's anomalies \cite{W1,W2}. In both cases the effect arises when the frequency approaches from the left to the frequency at which the new mode appears.

\begin{figure}[h!]
\centering
\includegraphics[width=0.4\textwidth,,height=0.25\textwidth]{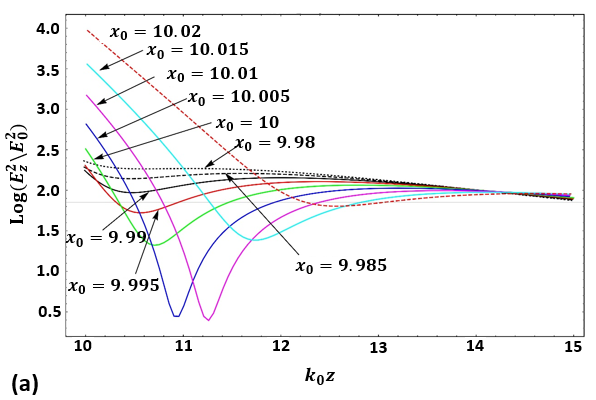}
\includegraphics[width=0.4\textwidth,,height=0.25\textwidth]{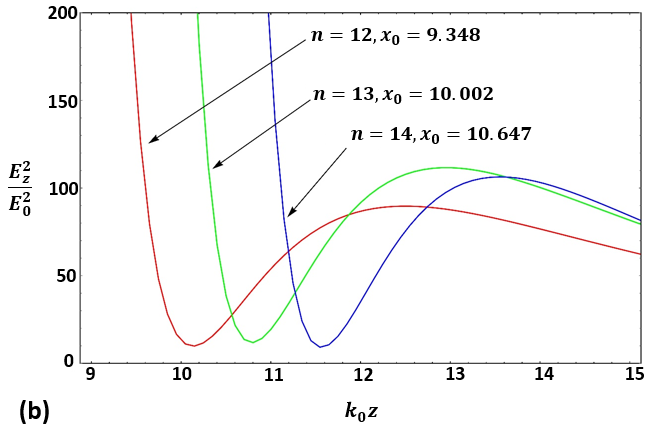}
\caption{(a). Coordinate dependencies of the normalized electric intensity on the beam axis behind the sphere for different 
size parameters of the sphere. (b). Similar dependencies for specific sizes corresponding to $n$-th mode resonance. In both plots $\varepsilon=3.1$.
}
\label{fig10}
\end{figure}

The distance from the sphere $\Delta z=z-a=\lambda/4$ at which the Fano minimum is observed for a near-field effect is substantial. Usually, near-field effects are observed either inside the resonant objects or at much smaller distances. In fact, there are parameters for which $\Delta z$ exceeds $\lambda/2$. Properties of our Fano resonance in a broad range of size parameters (for the same permittivity $\varepsilon=3.1$) are illustrated by Fig. \ref{fig10}. 

 {In Fig. \ref{fig10}(a) we see how the variation of the size parameter in the range $x_0 = 9.98-10.02$ results in the formation of several Fano resonances. In Fig. \ref{fig10}(b) we see that these resonant ranges are distant from one another by $\Delta x_0\approx 0.65$ over the axis of the optical sizes. So, our effect is strongly and periodically resonant.} 
 Similarly, for a given radius and frequency (chosen so that $a\gg\lambda$) there is a set of the resonant permittivity values. Finally, for a given permittivity and frequency, there is a set of the resonant values of $a$. The main Fano maximum always holds at the rear edge of the sphere, where the value of the normalized intensity $E_z^2/E_0^2$ is a huge value of the order $100-1000$. The intensity enhancement at this point depends on $x_0,\, \varepsilon$ and $\beta$. Also, the coordinate of the minimum and the normalized electric intensity at this minimum depend on these parameters. 

Notice that the near-field nature of our effect is seen not only from {Fig. \ref{fig9}} where we observe interference of two decaying exponential functions whose phases are nearly opposite. It is also seen from the field polarization at the Fano maximum and minimum and from the absence of the power transfer at these points. Exactly on the axis $z$ the power flux is identically zero (black arrows on the axis in Fig. \ref{fig5}(b) only show the direction of the Poynting vector in the paraxial region of the transmitted beam). Our effect is, though an interference, is not the conventional interference of two propagating waves of the same amplitudes which sum up at the points where their phases are the same and subtract at the points where they are opposite. 
On the contrary, it is the interference of two evanescent waves, whose phases are nearly opposite and the decay rate is different. The sharp minimum occurs at a point where the amplitudes of these evanescent waves equate one another. The spread maximum behind it results from the more rapid variation of the quasi-continuum phase. At the corresponding distances the evanescent component of the quasi-continuum decays sufficiently so that the propagating component would become important. This is the reason why the spatial variation of the phase in Fig. \ref{fig9} is more rapid for the quasi-continuum than for the resonant mode.  

\begin{figure}[h!]
\includegraphics[width=0.4\textwidth,,height=0.25\textwidth]{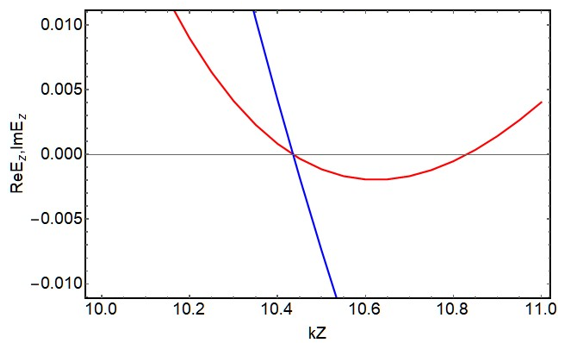}
\caption{The dependence of $\Re {E_Z}$ and $\Im {E_Z}$ on $k_0z$ for $\beta=0.00998899$, $x_0=9.356787479725$ and {$\varepsilon=3.1$}.}
\label{fig11}
\end{figure}

There is no lower limit for the minimal value of the electric field at the Fano minimum. In fact, in {Fig. \ref{fig5}(a)} one can see exact zero of the electric field. The dependence of ${E_Z}$ on $k_0z$ is shown in Fig. \ref{fig11}, where one can see that both real ($\Re$) and imaginary ($\Im$) parts of $E_z$ are equal to zero simultaneously.

{To exactly nullify} the electromagnetic field at the minimum of Fano resonance we should adjust the structure parameters. Two following equations should be satisfied:

{      
\begin{eqnarray}
\Re {E_z(k_0z,x_0,\beta,\varepsilon)}=0 \nonumber \\*
\Im {E_z(k_0z,x_0,\beta,\varepsilon)}=0 
\label{eq9}
\end{eqnarray}
}

\begin{figure}[h!]

\includegraphics[width=0.5\textwidth,,height=0.22\textwidth]{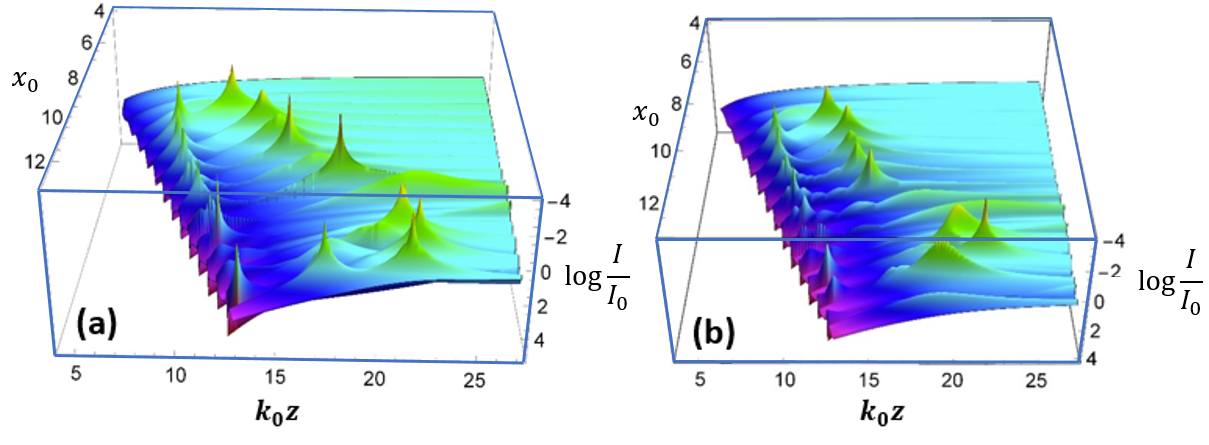}
\includegraphics[width=0.5\textwidth,,height=0.22\textwidth]{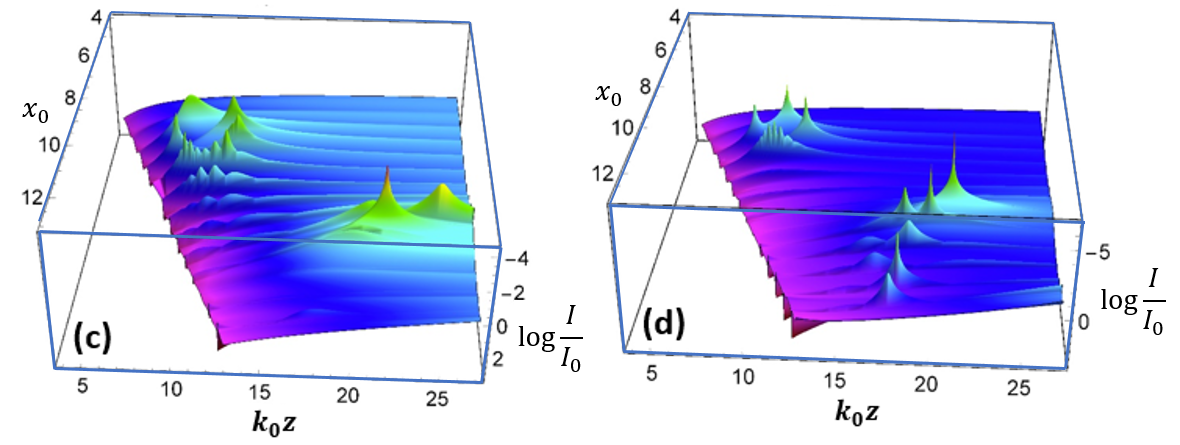}
\includegraphics[width=0.5\textwidth,,height=0.22\textwidth]{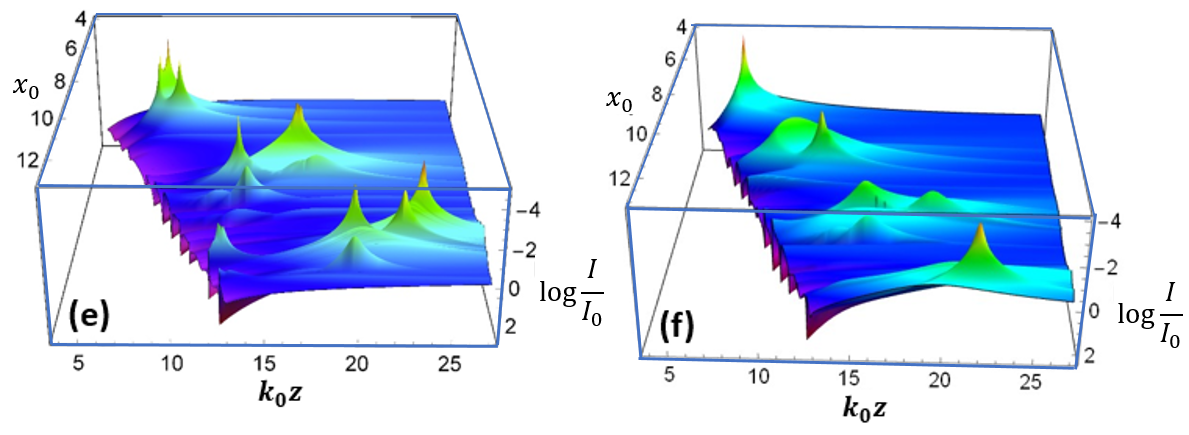}

\caption{Spatial distribution of the electric field intensity as a function of  the normalized coordinate $k_0z$ and size parameter $x_0$  for six conical angles $\beta$. Plots{(a)-(f) correspond to $\beta=0.01$, $\beta=0.157$, $\beta=0.25$, $\beta=0.314$, $\beta=0.452$, and $\beta=0.628$, respectively.}}
\label{fig12}
\end{figure}

\noindent
{Equation (\ref{eq9}) can hold at a certain point $k_0z$ behind the sphere due to the interplay between $x_0$, $\varepsilon$ and $\beta$. For any fixed parameter among these three, we can satisfy Eq. (\ref{eq9}) by tuning the other two.
It is difficult to finely tune $\varepsilon$ but we may generate the Bessel beam with any $\beta$ and adjust the frequency for 
any $x_0$.} We have also studied the behavior of our Fano resonance versus $\beta$. It is illustrated by Fig. \ref{fig12}.

From this figure one can see that for some modes (for $N=20$,12,10,7) the related Fano resonances cannot be seen because $\sin{N\beta}=0$ for these modes. The positions of minimum of Fano resonances are qualitatively the same. Another important point is decreasing of maximum field enhancement with increasing of $\beta$. This parameter has no noticeable impact until $\beta\approx\pi/2x_0$. This is the threshold value. If the conical angle of the Bessel beam is large, e.g. $\beta>\pi/k_0a$ it means that the sphere cross section covers two or more maxima of the incident beam intensity $I_i$ versus the cylindrical coordinate . Note, that in Fig. \ref{fig2}(b) we have sketched a near threshold situation, when the maximum of $I_i$ holds at $\rho\approx a$. In our reported calculations, it holds at $\rho\gg a$. So, smallness of $\beta$ is another condition of our effect together with $x_0\gg 1$. If $\beta$ is sufficiently small for whatever $\varepsilon> 2.5$ there is a set of optical sizes $x_0$ corresponding to our effect. All these values of $x_0$ are much larger than unity. For an experiment, one may take a popular optical glass SF11 whose permittivity in the red and orange bands of the visible range in accordance to the catalogues of Schott Inc. is equal to $\varepsilon=3.1$. Then in accordance to our {Figs. \ref{fig5}, \ref{fig6} and \ref{fig10}} for a microsphere with $a = 1.2\mu m$ we will obtain a set of resonant frequencies in the band of the red light, and with $a = 1.0\mu m$ -- in the orange light. This is only one example, there is a plenty of variants how to engineer this effect using available materials.

\section{Possible Applications of the Fano Spatial Resonance}
\label{sec3}
The revealed effect is, to our opinion, very promising for trapping the atoms and ions. An atom with polarizability $\alpha$, experiences in the monochromatic light of non-uniform intensity $I(x, y)$ the so-called gradient force

\begin{eqnarray}
F_g(x,y)=\frac{1}{2}\Re{(\alpha)}\nabla I(x,y)
\label{eq10}
\end{eqnarray}

This formula, initially derived in \cite{4} for dielectric nanoparticles, was generalized for atoms in the laser light field in \cite{4,5}. In the case of blue detuning (from the main excited state of the atom) $\Re{(\alpha)}<0$ and $F_g$ will be directed towards the minimum of the electric intensity \cite{balykin18}. Since our minima in Figs. \ref{fig3}, \ref{fig4} and \ref{fig5} are ultimately sharp, a trapped atom will be centered at the corresponding point of 3D space. 
Our singular Fano minimums can serve as a trap also for ions. In accordance with \cite{bialynicki20}, the paraxial region of a Bessel beam with radial polarization and nonzero order can serve a subwavelength thin trap for charged particles. It is possible to show that this trapping property keeps in our 3D case.
When the incident light has the same wavelength as that of the resonant optical absorption $\lambda_A$, or is slightly detuned (so that $\lambda$ is within the optical transition spectral line) and has sufficiently high flux density, the optical transition of an atom at $\lambda_A$ is pumped and the atom turns out to be cooled \cite{meystre21}. When the flux density is of the order of one $kW$ per square $cm$, optical potential of an atom expressed in Kelvins is of the order of dozens of $mK$. If the frequency detuning $\Delta=\omega-\omega_A$ of the incident light with respect to the transition frequency is positive, the polarizability of an atom has the negative real part and the trapping effect arises in the minimums of the electric intensity where the atom optical potential drops from dozens of $mK$ to one $mK$ or less \cite{sekatskii19,meystre21,letokhov22}.
Usually, the alkaline atoms are used for trapping. In Table \ref{tab} one can see main properties of such atoms \cite{alkaline}.
\begin{table*}[h!]

\caption{\label{tab}Main properties of alkaline atoms.}

\begin{tabular}{ | m {4em} | m{7em}| m{1.3cm} | m{2.3cm} | m{1.4cm} | m{3.3cm} |m{3.3cm} |} 
\hline
\hline

Atoms & Transition & $\lambda$ [nm] & Decay Rate $s^{-1}$ & Life time $nm$ & Transition Dipole
Matrix Element SI ($J=1/2||er||J'=3/2)$ & Transition Dipole Matrix Element SI ($J=1/2||er||J'=3/2)/\sqrt{2}$)\\ 
\hline

$^{133}Cs$ $D_2$ & $6^2S_{1/2}$→$6^2P_{3/2}$ & 852.347 & $3.281 \times10^7$ & 30.47 & $3.797 \times10^{-29}$ & $2.68 \times10^{-29}$ \\ 

$^{87}Rb$ $D_2$ & $5^2S_{1/2}$→$5^2P_{3/2}$ & 780.241 & $3.811 \times10^7$ & 26.24 & $3.584 \times10^{-29}$ & $2.53 \times10^{-29}$\\ 

$^{23}Na$ $D_2$ & $3^2S_{1/2}$→$3^2P_{3/2}$ & 589.158 & $6.154 \times10^7$ & 16.25 & $2.988 \times10^{-29}$ & $2.11 \times10^{-29}$ \\ 
\hline
\hline
\end{tabular}

\end{table*}

\begin{figure}[h!]

\includegraphics[width=0.4\textwidth,,height=0.3\textwidth]{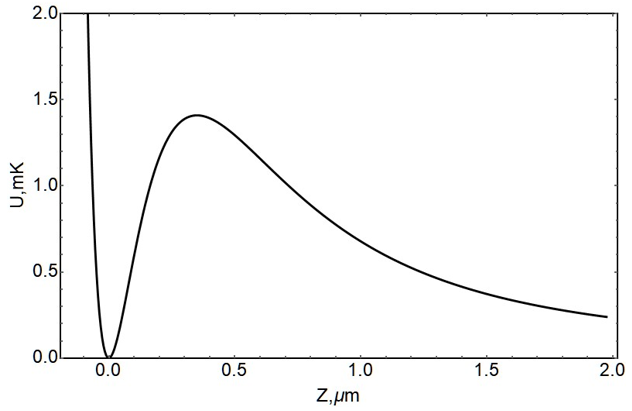}
\includegraphics[width=0.4\textwidth,,height=0.36\textwidth]{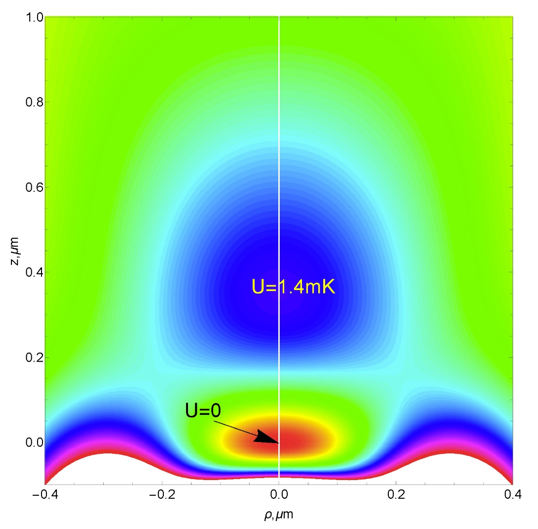}

\caption{The optical potential of an atom of $Cs$ for $\varepsilon=3.1$, $\beta= 0.009989$ and $x_0=9.3568$ 
(parameters offering $U=0$ at the trap center) varying along the beam axis (top panel), and versus both $\rho$ and $z$ (bottom panel). 
}
\label{fig13}
\end{figure}

All these atoms have similar properties and as an example Fig. \ref{fig13} depicts the optical potential $U(x,y)$ in our optical trap for the case when the trapped atoms are those of $Cs$ and the trap parameters are $\varepsilon=3.1$, $\beta= 0.01$ and $x_0=9.3567$. These parameters offer the zero to the potential at the trap center. We have calculated the distribution of $U$ for the wavelength of the incident beam $\lambda_A= 852nm$ when the atoms of Cs experience the transition of type $6S_{1/2}-6P_{3/2}\, D_2$. The trapping potential is calculated using the known formula (see e.g. in \cite{letokhov22}):

\begin{eqnarray}
U=\mu^2E^2/2k_B\hbar\Delta
\label{eq11}
\end{eqnarray}

\noindent
 where $\mu = 2.68\times10^{-29}$ SI is the dipole moment (matrix element) of the optical transition, 
 $k_B$ and $\hbar$ are Boltzmann and Planck constants, respectively, 
 $E^2 = I$ is the electric intensity, {corresponding to the power flux density $3 kW/cm^2$ in the intensity maximum of our incident beam}. In this case 
 we have $\Delta=1000\Gamma$, where $\Gamma=3.28\times10^7s^{-1}$ is the Lorentz damping frequency of the $Cs$ atom
 \cite{letokhov22}. 
{The laser pumping with the flux density $3 kW/cm^2$ and even higher one has been used for optical trapping since 1980s \cite{bialynicki20}}.

Approaching the minimum of the optical potential to zero, offers the long-time confinement of atoms, because the atoms located at the minimum practically do not heat up. As to the trap transverse sizes, in accordance to Fig. \ref{fig13} and the criterion of the trap bounds formulated in \cite{sekatskii19} (1 $mK$ above the potential minimal value) we have here the effective dimensions of our optical trap in the $z$ and $\rho$ directions nearly equal to 250 and 400 $nm$, respectively. So, our trap is subwavelength in the transverse plane ($0.46\lambda$) and strongly subwavelength ($0.29\lambda$) in the longitudinal direction. These sizes shrink one more if we slightly change the permittivity so that the minimal potential deviates from the exact zero. The similar shrink when the minimal potential becomes nonzero results from the 
tuning of the size parameter for the same permittivity $3.1$. The trap corresponding to $x_0=9.348,\, \beta=0.01$ depicted in Fig. \ref{fig6} is more subwavelength than that depicted in Fig. \ref{fig13}, as well as the trap corresponding to $x_0=10.641,\, \beta=0.01$ depicted in Fig. \ref{fig10}(b) (blue curve). In both these cases the trapping potential at the minimum is of the order of $1$ mK and the sizes shrink almost twice compared to the case when the potential nullifies at the trap center.

\section{Conclusion}

In this paper the general concept of spatial Fano resonance is put forward -- extended to the spatial domain. 
We have theoretically revealed a new near-field effect for optically large dielectric spheres illuminated by a Bessel beam with axial symmetry and slow oscillation in the transverse plane. This effect is a strong spatial Fano resonance for the longitudinal electric field that holds in an optically substantial region noticeably shifted behind the sphere. There are several papers studying the ordinary Fano resonances in the microsphere, e.g. \cite{Tribelsky24,kong25,wang26}, however, to our knowledge no one of these papers deals with the spatial Fano resonance outside the sphere and reveals the interference of evanescent waves. These Fano resonances hold over the frequency axis and imply a spectral hole neighboring the strong spectral maximum. Also, the Fano resonances over the frequency axis referred to the spectral minima and maxima of the internal electric or magnetic fields. Our effect is totally different. It can hardly be revealed for a microsphere impinged by a plane wave, where both $TM$ and $TE$ modes are excited simultaneously. When we illuminate our microsphere by a Bessel beam with the intensity profile similar to that shown in Fig. \ref{fig2}, the longitudinal electric field in the paraxial region dominates over the transverse field even in the incident beam. This determines the domination of the evanescent waves over the propagating waves in the transmitted field. And this domination is the prerequisite of the revealed effect. In this case the quasi-continuum of the non-resonant modes excited in the sphere may have the magnitude of the same order as that of the resonant mode. It enables their sharp interference and results in our effect. Unusual features of our effect is the interference of two exponentially decaying evanescent waves and the amazing sharpness of the Fano  of the electric intensity. The effect implies also the distant location of this intensity minimum (and the spread maximum located behind it) from the sphere. To our opinion, these two features are most interesting properties of our effect. 

We have found one work \cite{kim27} where a Bessel beam impinges a dielectric microsphere. However, in this work the authors considered a linearly polarized zero- order Bessel beam. This beam is not azimuthally symmetric and its electric field is polarized transversely on the beam axis (see e.g. in \cite{Mishra28,Chen29}). The excitation of evanescent waves in a microsphere by transversely polarized electromagnetic field is inefficient (see e.g. in \cite{Chen29}). Therefore, the authors of \cite{kim27} concentrated on the possibility to excite several photonic nanojets using a Bessel beam instead of a plane wave. In our case, we earn a strong near-field effect because the transverse electric field of the incident beam vanishes on the axis as it is shown in Fig. \ref{fig2}(a). In the paraxial region, the longitudinal electric field dominates. This feature of our incident beam enables the efficient generation of evanescent waves and results in the spatial Fano resonance. 
The existence of the spatial Fano resonance is proved here for the case of dielectric sphere illuminated by Bessel beam. However, the same effect can be observed in any symmetric geometry where only one field component is nonzero on the symmetry axis. As example of such geometry one can consider $TM$ and $TE$ plane waves normally incident on right circular cylinder \cite{klimov30} or spheroids illuminated by Bessel beams. 
For applications the importance of our effect is, probably, related to the rather high electric field gradient between the Fano maximum and Fano minimum behind the sphere. This secondary effect can be used for trapping the nanoparticles and maybe even atoms/molecules. For atoms/molecules we may adjust our trap so that it would grant the zero potential at the trap center (and very long trapping lifetime for cold atoms/molecules) or, alternatively, so that it would grant the trapping in a very small region (with the maximal size about 200 nm) if we allow the trapping potential to be of the order of 1 mK i.e. if the trapping lifetime is standard.  
Since the Bessel beams becomes a popular optical technique, a simple microsphere may be a promising and affordable tool for such the optical trapping.

\begin{backmatter}

\bmsection{Acknowledgments}  Funding by the Russian Foundation for Fundamental Investigations (grant №18-02-00315) is acknowledged by V.Klimov.

\bmsection{Disclosures} The authors declare no conflicts of interest.

\bmsection{Data Availability Statement} No data were generated or analyzed in the presented research.

\end{backmatter}

\bibliography{sample}

\begin{thebibliography}{10}
\newcommand{\enquote}[1]{``#1''}

\bibitem{1}
U.~D{\"u}rig, D.~W. Pohl, and F.~Rohner, \enquote{Near-field optical-scanning
  microscopy,} {\protect\JournalTitle{Journal of applied physics}} \textbf{59},
  3318--3327 (1986).

\bibitem{2}
H.~Yang, N.~Moullan, J.~Auwerx, and M.~A. Gijs, \enquote{Super-resolution
  biological microscopy using virtual imaging by a microsphere nanoscope,}
  {\protect\JournalTitle{Small}} \textbf{10}, 1712--1718 (2014).

\bibitem{3}
P.~K. Upputuri and M.~Pramanik, \enquote{Microsphere-aided optical microscopy
  and its applications for super-resolution imaging,}
  {\protect\JournalTitle{Optics Communications}} \textbf{404}, 32--41 (2017).

\bibitem{4}
A.~Ashkin, \enquote{Acceleration and trapping of particles by radiation
  pressure,} {\protect\JournalTitle{Physical review letters}} \textbf{24}, 156
  (1970).

\bibitem{5}
D.~S. Bradshaw and D.~L. Andrews, \enquote{Manipulating particles with light:
  radiation and gradient forces,} {\protect\JournalTitle{European Journal of
  Physics}} \textbf{38}, 034008 (2017).

\bibitem{6}
T.~Zacharias and A.~Bahabad, \enquote{Light beams with volume
  superoscillations,} {\protect\JournalTitle{Optics Letters}} \textbf{45},
  3482--3485 (2020).

\bibitem{limonov7}
M.~F. Limonov, M.~V. Rybin, A.~N. Poddubny, and Y.~S. Kivshar, \enquote{Fano
  resonances in photonics,} {\protect\JournalTitle{Nature Photonics}}
  \textbf{11}, 543 (2017).

\bibitem{fano8}
U.~Fano, \enquote{Effects of configuration interaction on intensities and phase
  shifts,} {\protect\JournalTitle{Physical Review}} \textbf{124}, 1866 (1961).

\bibitem{fano9}
U.~Fano, \enquote{Sullo spettro di assorbimento dei gas nobili presso il limite
  dello spettro d’arco,} {\protect\JournalTitle{Il Nuovo Cimento
  (1924-1942)}} \textbf{12}, 154--161 (1935).

\bibitem{connerade10}
J.-P. Connerade and A.~Lane, \enquote{Interacting resonances in atomic
  spectroscopy,} {\protect\JournalTitle{Reports on Progress in Physics}}
  \textbf{51}, 1439 (1988).

\bibitem{7}
D.~McGloin and K.~Dholakia, \enquote{Bessel beams: diffraction in a new light,}
  {\protect\JournalTitle{Contemporary Physics}} \textbf{46}, 15--28 (2005).

\bibitem{8}
V.~Klimov, \enquote{Manifestation of extremely high-q pseudo-modes in
  scattering of a bessel light beam by a sphere,} {\protect\JournalTitle{Optics
  Letters}} \textbf{45} (2020).

\bibitem{9}
A.~V. Osipov and S.~A. Tretyakov, \emph{Modern Electromagnetic Scattering
  Theory with Applications} (John Wiley Sons, Ltd, 2017).

\bibitem{W1}
R.~Wood, \enquote{On the remarkable case of uneven distribution of light in a
  diffraction grating spectrum,} {\protect\JournalTitle{Proceedings Phys. Soc.
  London}} \textbf{18}, 269--275 (1902).

\bibitem{W2}
L.~Rayleigh, \enquote{On the dynamical theory of gratings,}
  {\protect\JournalTitle{Proceedings Royal Soc. London A}} \textbf{9}, 399--416
  (1907).

\bibitem{balykin18}
V.~Balykin, V.~Klimov, and V.~Letokhov, \enquote{Handbook of theoretical and
  computational nanotechnology,atom nanooptics, eds., m. rieth and w.
  schommers,}  (2006).

\bibitem{bialynicki20}
I.~Bialynicki-Birula, Z.~Bialynicka-Birula, and N.~Drozd, \enquote{Trapping of
  charged particles by bessel beams,} {\protect\JournalTitle{The Angular
  Momentum of Light}}  (2012).

\bibitem{meystre21}
P.~Meystre, \emph{Atom optics}, vol.~33 (Springer Science \& Business Media,
  2001).

\bibitem{sekatskii19}
V.~Klimov, S.~Sekatskii, and G.~Dietler, \enquote{Laser nanotraps and
  nanotweezers for cold atoms: 3d gradient dipole force trap in the vicinity of
  scanning near-field optical microscope tip,} {\protect\JournalTitle{Optics
  communications}} \textbf{259}, 883--887 (2006).

\bibitem{letokhov22}
V.~Klimov and V.~Letokhov, \enquote{Laser focusing of cold atoms: Analytical
  solution of the problem,} {\protect\JournalTitle{Laser physics}} \textbf{13},
  339--349 (2003).

\bibitem{alkaline}
\enquote{Alkali d line data,}  (2021). \url{https://steck.us/alkalidata}.

\bibitem{Tribelsky24}
M.~I. Tribelsky, A.~E. Miroshnichenko, and Y.~S. Kivshar,
  \enquote{Unconventional fano resonances in light scattering by small
  particles,} {\protect\JournalTitle{Europhyiscs Letters}} \textbf{97} (2012).

\bibitem{kong25}
X.~Kong and G.~Xiao, \enquote{Fano resonance in high-permittivity dielectric
  spheres,} {\protect\JournalTitle{JOSA A}} \textbf{33}, 707--711 (2016).

\bibitem{wang26}
Z.~Wang, B.~Luk’yanchuk, L.~Yue, B.~Yan, J.~Monks, R.~Dhama, O.~V. Minin,
  I.~V. Minin, S.~Huang, and A.~A. Fedyanin, \enquote{High order fano
  resonances and giant magnetic fields in dielectric microspheres,}
  {\protect\JournalTitle{Scientific Reports}} \textbf{9}, 20293 (2019).

\bibitem{kim27}
M.-S. Kim, T.~Scharf, S.~M\"{u}hlig, C.~Rockstuhl, and H.~P. Herzig,
  \enquote{Engineering photonic nanojets,} {\protect\JournalTitle{Opt.
  Express}} \textbf{19}, 10206--10220 (2011).

\bibitem{Mishra28}
S.~R. Mishra, \enquote{A vector wave analysis of a bessel beam,}
  {\protect\JournalTitle{Optics Communications}} \textbf{85}, 159--161 (1991).

\bibitem{Chen29}
Y.~Chen, Z.~Han, Cui, and X.~Shi, \enquote{Scattering of a zero-order bessel
  beam by a concentric sphere,} {\protect\JournalTitle{Journal of Optics}}
  \textbf{16}, 055701 (2014).

\bibitem{klimov30}
V.~Klimov, R.~Heydarian, and C.~Simovski, \enquote{A dielectric microcylinder
  makes a nanocylindrical trap for atoms and ions,}
  {\protect\JournalTitle{arXiv preprint arXiv:2012.04301}}  (2020).

\end{thebibliography}



\ifthenelse{\equal{\journalref}{aop}}{%
\section*{Author Biographies}
\begingroup
\setlength\intextsep{0pt}
\begin{minipage}[t][6.3cm][t]{1.0\textwidth} 
  \begin{wrapfigure}{L}{0.25\textwidth}
    \includegraphics[width=0.25\textwidth]{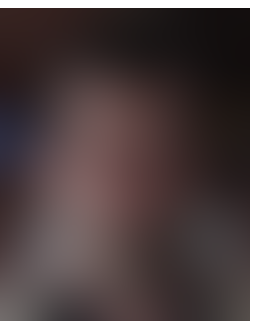}
  \end{wrapfigure}
  \noindent
  {\bfseries John Smith} received his BSc (Mathematics) in 2000 from The University of Maryland. His research interests include lasers and optics.
\end{minipage}
\begin{minipage}{1.0\textwidth}
  \begin{wrapfigure}{L}{0.25\textwidth}
    \includegraphics[width=0.25\textwidth]{alice_smith.eps}
  \end{wrapfigure}
  \noindent
  {\bfseries Alice Smith} also received her BSc (Mathematics) in 2000 from The University of Maryland. Her research interests also include lasers and optics.
\end{minipage}
\endgroup
}{}

\end{document}